**Accepted manuscript — forthcoming in *Synthese***

# Distributional Counterfactual Dependence in the Special Sciences

Youheng Zhang
School of Marxism, University of Electronic Science and Technology of China, Chengdu, Sichuan, China
Corresponding author: Youheng Zhang; email: zhangyouheng@whut.edu.cn
ORCID: https://orcid.org/0000-0002-2363-487X

**Abstract:** Special-science explanations often concern distributions rather than individual outcomes. Since higher-level states depend on lower-level processes, what would make a higher-level explanation redundant? I argue that redundancy requires replacement without loss, not merely a reconstruction of individual outcomes. Distributional counterfactual dependence (DCD) identifies how changes in higher-level conditions alter an outcome distribution or a relevant feature of it. Against the background of stochastic supervenience, I distinguish this explanatory role from the dependence of higher-level states on lower-level bases. A replacement must be specified independently of the distribution to be explained and preserve the same counterfactual pattern at the grain and across the conditions relevant to the question. Applications to wage distributions, cognitive load, and selection clarify this requirement. A hypothetical comparison shows how a lower-level model can preserve the explanation and why matching selected outcome statistics may not suffice. DCD thus describes one form of non-reductive explanation that depends on counterfactual organization rather than additional higher-level productive powers.

**Keywords:** Counterfactual dependence; counterfactual explanation; special sciences; explanatory autonomy; stochastic supervenience

## 1 Introduction

The debate over the autonomy of the special sciences is less a dispute about disciplinary classification or theoretical reduction than a metaphysical question about whether higher-level explanations possess genuine explanatory standing[1]. If higher-level concepts in psychology, biology, or the social sciences are merely convenient summaries of lower-level physical facts, their claim to non-redundant explanatory standing becomes difficult to sustain, even if they facilitate prediction and categorization.

A familiar anti-reductionist response appeals to multiple realizability: the idea that the same higher-level property can be realized by different physical states, so that the classifications of the special sciences need not map onto those of physics (Fodor, 1974, 1997). But multiple realizability at most supports the irreducibility of higher-level classifications; it does not by itself secure their unified explanatory role (Shapiro, 2000; Sober, 1999). As long as higher-level properties depend on their physical realizers, and the relevant outcomes can also be characterized in lower-level terms, their non-redundant explanatory role remains under pressure (Bennett, 2003; Meyering, 2000; Raatikainen, 2010).

Counterfactual theories offer a more modest account of higher-level explanation by treating explanatory relevance, and often causal relevance, as difference-making rather than as the exercise of additional productive powers (Lewis, 1973, 2000). In the special sciences, such difference-making often concerns patterns rather than dependencies between individual events: changes in mental load, interest rates, selection pressures, or institutions alter outcome probabilities without fixing particular outcomes. Interventionist, invariance, and realization-insensitivity accounts emphasize stable patterns of this kind (Menzies & List, 2010; Woodward, 2000), while work on probabilistic, exception-tolerant generalizations treats them as legitimate forms of counterfactual explanation (Reutlinger, 2013; Sandgren & Steele, 2021; Schrenk, 2006).

[1] Throughout, "higher-level" and "lower-level" are shorthand for inquiry-relative descriptions linked by a specified realization, composition, or coarse-graining relation. The terms do not imply that these relations align in a single global hierarchy; Potochnik (2021) argues that they often come apart and that their relevance is local to particular inquiries.

The counterfactual approach, however, cannot bypass Kim's challenge. Counterfactual conditionals can mark genuine relations of dependence, but such relations need not be causal or explanatorily autonomous (Kim, 1973, 1974). Once higher-level properties are understood as dependent on lower-level realizers, mere higher-level difference-making is not enough to show that the higher-level explanation is non-redundant (Kim, 1992, 1995, 2003). The metaphysical question raised by event-level counterfactual dependence is whether the dependence it captures can be anything more than a summary of lower-level facts. Answering this question requires specifying what kind of dependence higher-level explanations must exhibit if they are to remain genuinely non-redundant under lower-level dependence.

I argue that Kim's challenge to higher-level explanation in the special sciences should be recast as a problem of replacement without loss. The missing standard is not one more test of higher-level relevance, but a criterion for when lower-level reconstruction preserves the explanation it purports to replace. Distributional counterfactual dependence (DCD) supplies a standard for these cases: redundancy requires an independently specified lower-level reconstruction that preserves the explanandum and contrast, the relevant distributional feature, the background range, and the explanatory grain. Against the background of stochastic supervenience, DCD characterizes how higher-level conditions organize counterfactual outcome distributions without positing productive powers beyond their lower-level bases (cf. Kim, 2006, 2013; see Zhang, 2026).

Section 2 reconstructs Kim's challenge and introduces stochastic supervenience as the background relation between lower-level bases and higher-level properties. Section 3 develops DCD and applies it to three special-science cases. Section 4 situates DCD among nearby accounts of counterfactual, probabilistic, and distributional explanation, and explains why lower-level recoverability is not yet explanatory replacement.

## 2 Kim's Challenge and Stochastic Supervenience

The problem with counterfactual dependence is not that it fails to register dependence, but that it registers it too thinly. A counterfactual conditional can be true because one fact is conceptually, institutionally, or metaphysically determined by another, rather than because the former is causally produced by the latter. Counterfactual dependence can therefore mark a pattern of dependence, but it does not by itself tell us what kind of dependence this is, nor whether it involves the generative or productive connection required for explanation. This is the core of Kim's critique of counterfactual analyses (Kim, 1973), and it is reinforced in his later distinction between dependency causation and productive causation (Kim, 2023).

The case of Socrates and Xanthippe illustrates the point. Given their marital relation, Socrates' death at a given time makes Xanthippe a widow at that time. Hence the counterfactual "Had Socrates not died at that time, Xanthippe would not have become a widow at that time" may well be true. Yet this is not causation in the ordinary sense. Xanthippe's becoming a widow is not produced by Socrates' death through a causal process; it is non-causally determined by that death, together with the marital relation and the relevant conceptual conditions (Kim, 1974). The example shows why the kind of dependence matters.

The same limitation applies to higher-level counterfactuals in the special sciences. Conditions such as cognitive load, market expectations, selection pressures, and institutional arrangements can support genuine counterfactual judgments, but that is insufficient to establish the non-redundancy of higher-level explanations (e.g., Hédoin, 2019; Muhlert, 2025; Oulasvirta & Hornbæk, 2022; Qian, 2025). The issue is whether such covariation constitutes a non-redundant explanatory pattern rather than a dependence pattern fixed by lower-level realizers.

Many special-science explananda concern distributional patterns rather than individual events: error rates, reaction times, risks, investment tendencies, frequencies, and similar quantities (Woodward, 2000). In such cases, event-level counterfactual dependence does not by itself capture the explanandum. The question is whether the probability structure of the outcome variable would systematically differ under different higher-level conditions[2]. But the distributional turn alone does not secure autonomy. A higher-level distributional dependence can still be derivative of lower-level bases, so the non-redundant role of the higher-level classification remains to be established (Kim, 1992, 2003, 2007).

The explanatory task is therefore to situate higher-level counterfactuals within a metaphysical account of dependence. Counterfactual conditionals register covariation, whereas supervenience specifies how higher-level facts depend on lower-level bases. The question is whether this dependence can constrain higher-level outcomes without requiring every lower-level base to fix a unique higher-level state (cf. Kallestrup, 2006; Kim, 2010, 2014; Raatikainen, 2010).

Stochastic supervenience provides such an account. Informally, a lower-level base, together with the relevant laws or background conditions, fixes an objective probability distribution over higher-level states without fixing which state occurs. The residual variation is not mere ignorance: different admissible bases lawfully determine their associated probability measures. Deterministic supervenience is the limiting case in which the relevant measure assigns probability one to a single higher-level state. Stochastic supervenience therefore specifies the metaphysical dependence relation without yet determining which features of the resulting distribution matter to the explanandum.

Formally, a higher-level property space $A$ stochastically supervenes on a lower-level base space $B$, relative to a background set of laws or constraints $L$, when each admissible lower-level base state $b \in B$ lawfully fixes not a unique higher-level property, but a probability measure over higher-level properties. I previously represented this dependence by a Markov kernel (Zhang, 2026):

$$\Phi: B \to \Delta(A).$$

For each $b \in B$, $\Phi(b)$ is a probability measure over $A$. The dependence is supervenience-like in that the lower-level base, together with the laws, fixes the relevant probability measure; it is stochastic in that what gets fixed is a distribution, not a single higher-level value. My full account adds further constraints to block trivial relabeling, mere statistical association, and ignorance-based interpretations of the residual stochasticity.

Deterministic supervenience is the limiting case. If $f: B \to A$ assigns each lower-level base a unique higher-level property, then:

$$\Phi(b) = \delta_{f(b)},$$

where $\delta_{f(b)}$ is the Dirac measure concentrated on $f(b)$. Stochastic supervenience therefore generalizes classical supervenience rather than rejecting it: higher-level dependence consists in the stable constraint of probability distributions rather than the determination of single outcomes.

## 3 Distributional Counterfactual Dependence

### 3.1 Formalizing DCD

Against the background of stochastic supervenience, counterfactual dependence in the special sciences can be expressed as a counterfactual difference between outcome distributions. This formulation applies only where the distribution is part of the explanandum, not merely a device for summarizing ignorance or measurement error. In such cases, variation in the outcome distribution, or in some feature of it, is the structure to be explained.

[2] Appealing to probability structures here does not imply that every modeled distribution has metaphysical significance. Some probability assignments reflect measurement limits, modeling choices, or incomplete information; more detailed lower-level information may support deterministic predictions instead (Bacciagaluppi, 2020; Lyon, 2011; Pietsch, 2022). The argument concerns cases in which, relative to the base partition fixed by the inquiry and the relevant background laws, the outcome distribution is itself a stable structure to be explained (see Hoefer, 2007; List & Pivato, 2021). Stability alone does not establish objectivity. In Zhang (2026), I set out the further conditions under which such a distribution represents an objective constraint imposed by lower-level bases rather than merely an epistemic device.

Let $X$ be a higher-level explanatory variable with admissible value space $\mathcal{X}$, and let $Y$ be an outcome variable with outcome space $\mathcal{Y}$. Values of $X$ represent the higher-level conditions relevant to the explanation. For each admissible higher-level condition $x \in \mathcal{X}$, let $Y_x$ denote the outcome variable under the counterfactual supposition that $X = x$[3]. This counterfactual variable induces a probability measure on $\mathcal{Y}$:

$$\mu_x^Y = \mathcal{L}(Y_x) \in \Delta(\mathcal{Y}).$$

Here, $\mathcal{L}(Y_x)$ denotes the law of $Y_x$, and $\Delta(\mathcal{Y})$ denotes the set of probability measures on $\mathcal{Y}$. For any measurable outcome set $E \subseteq \mathcal{Y}$,

$$\mu_x^Y(E) = P(Y_x \in E).$$

The outcome variable $Y$ exhibits minimal DCD with respect to the higher-level variable $X$ iff there exist admissible values $x_1, x_2 \in \mathcal{X}$ such that the counterfactual laws of $Y$ under those conditions differ:

$$\exists x_1, x_2 \in \mathcal{X} \text{ such that } \mu_{x_1}^Y \neq \mu_{x_2}^Y.$$

That is, two admissible higher-level conditions induce distinct counterfactual outcome distributions for $Y$. Equivalently, the distributions differ on at least one measurable subset of $\mathcal{Y}$.

The same relation can be represented by the distribution-valued mapping:

$$\Psi_Y : \mathcal{X} \to \Delta(\mathcal{Y}),$$
$$\Psi_Y(x) = \mu_x^Y = \mathcal{L}(Y_x).$$

On this representation, minimal DCD consists in the non-triviality of $\Psi_Y$:
for some admissible $x_1, x_2 \in \mathcal{X}$,

$$\Psi_Y(x_1) \neq \Psi_Y(x_2).$$

The mapping $\Psi_Y$ should be distinguished from the stochastic-supervenience mapping $\Phi$. The subscript on $\Psi_Y$ marks that DCD is outcome-specific: different explananda may induce different counterfactual mappings, whereas $\Phi$ names the background supervenience relation between lower-level bases and the higher-level property space. These two mappings are formally similar in that both are distribution-valued dependence relations, but they enter the argument at different points. $\Phi$ describes how lower-level bases probabilistically constrain higher-level properties. $\Psi_Y$, by contrast, describes how higher-level conditions organize counterfactual outcome distributions. Thus $\Phi$ supplies the metaphysical background of dependence, while $\Psi_Y$ captures the counterfactual organization of the explanandum. This distinction does not give higher-level conditions an autonomous force detached from lower-level bases. The question is instead whether a higher-level classification corresponds to a systematic counterfactual difference in outcome distributions, rather than merely to an epistemic classification or observational association. The contrast is summarized in **Fig. 1**.

[3] I use $Y_x$ as a compact counterfactual-variable notation, abstracting from potential-outcome notations such as Rubin's $y(E)$ and $y(C)$ and Holland's $Y_t$ and $Y_c$ (Holland, 1986; Rubin, 1974). The contrast between $Y_x$ and $Y \mid X = x$ follows the standard distinction between counterfactual or interventional quantities and observational conditionals (Pearl, 2009a, 2009b).

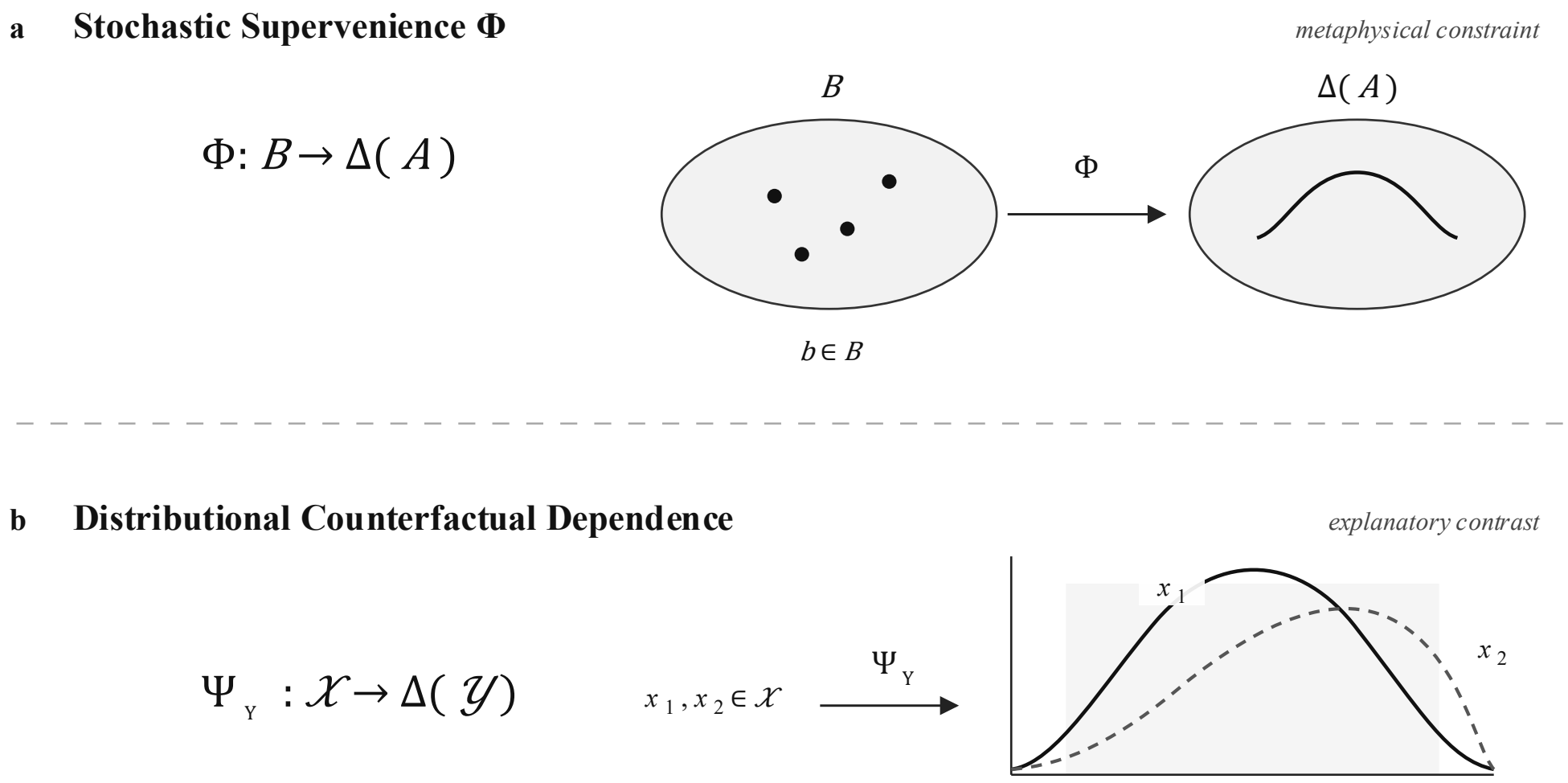

**Fig. 1 Two distribution-valued dependence relations.**

This distinction guides the cognitive-load application in subsection 3.2.2. Stochastic supervenience concerns how neural and physical bases constrain probabilities over psychological load or control states, whereas DCD concerns how experimentally varied load conditions change the joint distribution of accuracy and response time. The former specifies the dependence of psychological states on their bases; the latter identifies the counterfactual pattern that a lower-level account would have to preserve.

Minimal DCD is not sufficient for explanatory significance. A mathematically real difference between probability measures may be explanatorily idle if it concerns no relevant feature of the explanandum, depends on an arbitrary modeling choice, or disappears under reasonable variation in background conditions. Explanatory DCD therefore requires a distributional difference tied to the explanatory contrast, relevant background range, and grain.

The final requirement, explanatory grain, is question-relative. It specifies which differences among conditions and outcomes matter to the explanation and which may be ignored. Two accounts preserve the same grain when they answer the same contrastive question, track the same distributional feature, and support the same type of generalization across the admissible background range. An event-level account may contain more detail yet answer a different question (cf. Potochnik, 2010).

Let $\mathcal{C}$ denote the space of background conditions, and let $\mathcal{C}^* \subseteq \mathcal{C}$ denote the relevant background range fixed by the explanatory context, such as an experimental paradigm, institutional setting, ecological environment, or set of modeling assumptions[4]. The range $\mathcal{C}^*$ is neither the set of all metaphysically possible backgrounds nor a post hoc restriction to the observed case; it is the range presupposed by the explanatory practice at issue. For $c \in \mathcal{C}^*$, write:

$$\mu^Y_{x,c} = \mathcal{L}(Y_{x,c}).$$

Let $T$ be the distributional functional or feature relevant to the explanandum, such as a mean, variance, error rate, risk, quantile, frequency, tail probability, inequality measure, or structural feature. The choice of $T$ is not arbitrary: it is fixed by the explanatory question and by the standards of the relevant special science, not selected after the fact to generate a contrast. Explanatorily significant DCD requires the contrast between $x_1$ and $x_2$ to register in $T$ across the relevant background range. A strong uniform version of this requirement can be represented as:

$$\exists x_1, x_2 \in \mathcal{X} \text{ such that } \forall c \in \mathcal{C}^*,\ T(\mu^Y_{x_1,c}) \neq T(\mu^Y_{x_2,c}).$$

[4] The relevant background range does not fix all lower-level details or require persistence under arbitrary possible changes. It consists of the conditions presupposed by a given special-science explanation. Stability requires preservation across reasonable variations within that range. On stability and invariance in explanation, see Woodward (2004).

The explanatory structure at issue may be more specific than mere inequality, for example a fixed direction of change, an ordering among conditions, or another stable pattern of variation across $\mathcal{C}^*$.

Two contrasts are important. First, DCD differs from event-level counterfactual dependence: it concerns changes in counterfactual outcome distributions, or in explanatorily relevant features of those distributions, rather than the occurrence or non-occurrence of particular events. Second, DCD differs from observational association. Observational association compares actual conditional distributions, such as:

$$\mathcal{L}(Y \mid X = x_1) \text{ and } \mathcal{L}(Y \mid X = x_2).$$

DCD instead compares the laws of counterfactual variables:

$$\mu^Y_{x_1} = \mathcal{L}(Y_{x_1}) \text{ and } \mu^Y_{x_2} = \mathcal{L}(Y_{x_2}).$$

It expresses a counterfactual difference between outcome distributions, not a statistical difference between observed groups. This still leaves the exclusion question open: can the distributional role induced by a higher-level condition be preserved by a lower-level description?

## 3.2 Applications

The three applications use the same test. A lower-level account must be specified independently of the distribution to be explained. Under warranted assumptions relating the lower- and higher-level variables, it must preserve the explanandum and contrast, the relevant distributional feature, the background range, and the explanatory grain. Reproducing the observed distribution or explaining individual outcomes does not by itself meet this test. Table 1 previews the comparison.

**Table 1. The lossless-replacement test across the three applications**

| Case | Explanandum and contrast | Insufficient lower-level account | Lossless replacement retains |
|---|---|---|---|
| **Wage distributions** | Institutional effects on quantiles, lower-tail inequality, or wage compression | Individual-earnings explanation or observed-distribution reconstruction | Institutional contrast<br>Relevant functional<br>Labor-market range<br>Institutional grain |
| **Cognitive load** | High/low-load shift in the joint accuracy–response-time distribution | Participant-level fit or individual-response explanation | Load contrast<br>Joint distributional feature<br>Task/participant range<br>Psychological grain |
| **Selection** | Trait- or allele-frequency change across regimes | Birth/death reconstruction or a single observed frequency path | Regime contrast<br>Frequency functional<br>Population/environment range<br>Population grain |

### 3.2.1 Wage Distributions

Labor economics offers a clear case because many central explanations target wage distributions rather than individual wage outcomes. Minimum wages, unionization, and wage-setting institutions help generate those distributions. Studies of wage inequality therefore ask how these conditions affect quantiles, quantile gaps, low-wage shares, variances, or inequality indices, rather than how a particular worker's wage is determined (DiNardo et al., 1996). Let $X$ denote labor-market institutions and $Y$ wages; the explanandum is the counterfactual wage distribution or its relevant functionals.

Distributional methods in labor economics make this framing concrete. Fortin et al. (2011) separate composition effects from changes in the wage structure. Their framework extends decomposition beyond mean wages to quantiles, variances, and inequality measures. Chernozhukov et al. (2013)

provide a more general framework for estimating counterfactual distributions and functionals such as quantile functions, Lorenz curves, and Gini coefficients.

These methods do not identify the observed conditional distribution $\mathcal{L}(Y \mid X = x)$ with the counterfactual distribution $\mathcal{L}(Y_x)$. They construct counterfactual distributions under reweighting procedures, support conditions, and identification assumptions, which help fix the relevant background range $C^*$.

Under the DCD framework, explanations invoking labor-market institutions target the counterfactual wage distributions those institutions induce, rather than individual earnings directly. Let $x_{\text{high}}$ and $x_{\text{low}}$ stand for an institutionally relevant contrast, such as higher versus lower real minimum wages, or higher versus lower union density. If the object of inquiry is an economically significant distributional functional $T_q$, such as a quantile, a quantile gap, or an inequality index, the relevant contrast can be written as:

$$T_q(\mu^Y_{x_{\text{high}},c}) \neq T_q(\mu^Y_{x_{\text{low}},c}),$$

for background conditions $c$ within the institutionally relevant range $C^*$. The claim is not that the entire wage distribution must differ in every respect, but that the institutional contrast is preserved in features such as lower-tail shifts, wage compression, quantile gaps, or inequality measures.

Lower-level detail remains relevant. Individual skills, firm decisions, bargaining, mobility, matching, and shocks can improve mechanism specification and local prediction. But if a lower-level account reproduces the distribution only by listing heterogeneous worker-, firm-, and transaction-level facts, without preserving the institutional organization of lower-tail shifts, wage compression, or quantile changes, it provides a micro-level reconstruction rather than a lossless replacement. A shift from distributional change to why a particular worker received a particular wage changes the explanandum.

### 3.2.2 Cognitive Load

In experimental psychology, manipulations of cognitive load, working-memory load, or executive control specify task conditions under which patterns of performance are expected to change. Sweller (1988) treats cognitive load as the demand that problem-solving activity places on limited cognitive processing capacity, arguing that means-ends analysis can consume resources that would otherwise support schema acquisition. Engle (2002) characterizes working-memory capacity as closely connected to executive attention, especially in conditions where interference or competing response tendencies must be controlled.

Here $X$ denotes the relevant load or control condition, and $Y$ denotes response time, accuracy, or the pair of response time and accuracy. The diffusion decision model, for example, is designed to account jointly for accuracy, mean response times, and the shape of response-time distributions for both correct and error responses (Ratcliff & McKoon, 2008).

In this context, DCD takes the form of a contrast between performance distributions under different task demands. Let $x_{\text{high}}$ and $x_{\text{low}}$ denote high and low cognitive-load conditions. Cognitive load is explanatorily relevant insofar as it systematically shifts performance-relevant functionals of the distribution:

$$T_{\text{perf}}(\mu^Y_{x_{\text{high}},c}) \neq T_{\text{perf}}(\mu^Y_{x_{\text{low}},c}),$$

for background conditions $c$ within the range $C^*$ fixed by the experimental paradigm. Here $T_{\text{perf}}$ may extract error rate, mean latency, variance, tail behavior, or a feature of the joint accuracy–response-time distribution.

Experiments can vary memory load while controlling stimulus presentation and response format. Rescorla's interventionist account explains the relevance of such manipulations: psychological explanations specify how outcomes would change under interventions on psychological or sensory conditions (Rescorla, 2018). The background range $C^*$ is fixed by task type, stimuli, response format, participant population, and measurement procedure. The replacement test asks whether a neural model explains the same performance contrast across $C^*$.

Consider a hypothetical two-choice task in which responses are classified as fast or slow. The explanandum is an increase in slow errors under high rather than low load, across two stimulus difficulties in a fixed participant population. In a simple neural transmission model, successful transmission yields a fast, correct response; failure yields a slow guess with equal chances of being correct or incorrect. The physical demands of a concurrent task fix the load contrast. Transmission probabilities are 0.8 under low load and 0.6 under high load for easier stimuli, and 0.6 and 0.4 for harder stimuli. Assume that evidence independent of the target behavioral distributions supports these probabilities and the rule linking transmission to responses.

The probability of a slow error is half the probability of transmission failure. It therefore rises from 0.1 to 0.2 for easier stimuli and from 0.2 to 0.3 for harder stimuli. The rule also determines the other joint outcomes. Under these assumptions, replacement is lossless for the specified question because the model derives the same load-related increase in slow errors across both difficulties, at the stipulated fast/slow grain. Differences among neural realizations are irrelevant to this explanation when they leave the transmission probabilities unchanged.

Now consider a second lower-level reconstruction with the same error rate and fast-response rate in each condition. Here, errors arise during fast neural readout, while slow responses allow correction. For easier stimuli under low load, it assigns probabilities of 0.7 to fast correct responses, 0.1 to fast errors, and 0.2 to slow correct responses. The error rate is still 0.1 and the fast-response rate 0.8, but slow errors have zero probability under either load condition. Matching the two marginal distributions thus leaves the load-related contrast in their joint distribution unexplained. It could answer a narrower question about overall error rates, but not the question about slow errors.

Ma et al. (2026) manipulated cognitive load in a pedestrian-crossing task and fitted drift–diffusion models to participants' choices and response times. The best-fitting model treated cognitive load as reducing responsiveness to perceptual evidence. Its evidence trajectories were compared with centro-parietal positivity in the EEG record. The comparison makes the relation between the behavioral contrast and the neural measure explicit, but it does not establish lossless replacement.

### 3.2.3 Selection

Selection explanations concern how trait frequencies and reproductive success would vary under different ecological conditions. The first question is whether the proposed selection pressure supports such a counterfactual contrast. Inferring a pressure solely from the frequency change it is meant to explain would merely rename the outcome. Abrams (2006) explains how claims about counterfactual or idealized populations can support claims about population-level probabilities. The relevant background range $C^*$ must be fixed by ecological conditions, genetic assumptions, population structure, model idealizations, and competing evolutionary factors.

In this case, $X$ denotes a selection environment or selection pressure, while $Y$ denotes traits, fitness, reproductive outcomes, or offspring number. The explanandum is not whether a particular organism survives or reproduces, but whether trait frequencies, fitness distributions, mean reproductive success, or population-level trait distributions change. This framing fits the philosophical debate over selection. Millstein (2006) treats selection as a population-level causal process; Glennan (2009) separates individual-level causal productivity from population-level causal relevance; Huneman (2012) develops a counterfactual account of selection's causal relevance without adding a distinct productive force.

DCD represents this relevance as a counterfactual difference in a population-level distribution. Let $x_1$ and $x_2$ represent distinct selective environments, such as different predation regimes, resource environments, or population compositions. If the explanatory focus is trait frequency, the relevant contrast can be written as:

$$T_{\text{freq}}(\mu^Y_{x_1,c}) \neq T_{\text{freq}}(\mu^Y_{x_2,c}),$$

for background conditions $c$ within the ecologically and genetically relevant range $C^*$. Once the selective contrast has independent support, replacement requires an individual-level model that derives the same distributional change across $C^*$. Reconstructing a sequence of births and deaths is insufficient if it leaves the consequences of alternative environments unspecified. This requirement is compatible with the statisticalist denial that selection pressure is a productive cause distinct from individual processes.

Frequency-dependent selection shows what a reconstruction can lose. For example, predation on a color form can depend on how common it is (Glennan, 2009). A model might reproduce survival at one observed population composition yet fail to explain how fitness changes as that composition changes. Holding the observed survival rate fixed would miss this dependence. An individual-level model of predator encounters could instead derive the change across population compositions, treating population structure as a background condition.

The biological case therefore separates two questions: whether a selective contrast has independent support, and whether a lower-level model preserves its explanatory role across the same ecological range. The use of population-level variables alone answers neither.

## 4 Discussion

### 4.1 Counterfactual Dependence

DCD preserves the difference-making intuition while shifting its object from individual events to counterfactual outcome distributions, or their explanatorily relevant features. Lewis's analysis gives the classical event-level formulation: event $c$ causes event $e$ when, roughly, had $c$ not occurred, $e$ would not have occurred (Lewis, 1973, 1979). But, as Davis (1980) notes, counterfactual dependence can occur without causation, causation can occur without straightforward counterfactual dependence, and causation is asymmetric in a way counterfactual dependence need not be. The issue is not whether counterfactuals matter, but how the relevant dependence relation is structured.

Kim's exclusion argument gives this problem its metaphysical force: if higher-level properties depend on lower-level bases, higher-level counterfactual covariance does not by itself establish non-redundant explanation (Kim, 1992, 1995, 2003). The question is whether higher-level counterfactuals capture the dependence relation at the right explanatory grain.

In its explanatorily significant form, DCD brings together the counterfactual idea of difference-making with two further constraints: proportionality and stability. Yablo's proportionality constraint shows why more specific lower-level descriptions are not automatically better explanations (Yablo, 1992), while Woodward's interventionist account ties explanatory relevance to stable and invariant relations among variables (Woodward, 2000, 2021a). DCD applies these constraints to distributional explananda by treating higher-level conditions as inducing counterfactual outcome distributions whose relevant features must be stable across the background range fixed by the explanation, in the sense formalized by $\Psi_Y$ in Section 3. This does not settle the autonomy question; it specifies the distributional form that counterfactual dependence must take when the explanandum is a stable pattern in outcomes.

This distributional focus leaves room for non-reductive explanations of particular events. When the explanandum is a particular event, a proportionality account of token causation may identify a higher-level state as the proportionate, realization-insensitive difference-maker (Menzies & List, 2010). A distributional explanandum sets a different standard: an account of individual outcomes is a lossless replacement only if it also preserves their organization across the relevant conditions.

### 4.2 Probabilistic Counterfactuals

DCD is adjacent to work on probabilistic and chance-based counterfactuals, but its target differs. Probabilistic counterfactuals ask what probability an event or proposition would have under a counterfactual supposition; DCD asks whether varying a higher-level condition changes an outcome distribution or its explanatorily relevant features.

In variable-based notation, a probabilistic counterfactual can be represented as a claim such as:

$$P(Y_x \in E) = p,$$

where $E \subseteq \mathcal{Y}$ is an outcome event and $Y_x$ denotes the outcome variable under the supposition that $X = x$. Related causal-modeling approaches evaluate counterfactual or interventional probabilities such as:

$$P(Y \in E \mid do(X = x)) = p.$$

This terrain has been explored along several lines: probabilistic and chance-based counterfactuals (Barker, 1999; Finch, 1958; Leitgeb, 2012; Santorio, 2025); the relation between conditionals and probability (Hájek, 2003, 2011, 2012, 2014, 2021a, 2021b, 2025); and causal-modeling approaches to specific counterfactual probabilities (Balke & Pearl, 1994; Pearl, 1999).

By contrast, DCD does not ask merely what probability a given event would have under one counterfactual or interventional supposition. At the minimal level, it asks whether different higher-level conditions induce different counterfactual outcome distributions:

$$\mu^Y_{x_1} \neq \mu^Y_{x_2}.$$

Where the explanatory target is a particular distributional feature, the relevant contrast is instead:

$$T(\mu^Y_{x_1,c}) \neq T(\mu^Y_{x_2,c}),$$

relative to the background range fixed by the explanatory context. The relevant object is not a single probabilistic counterfactual claim, but a distribution-valued dependence relation plus restrictions on relevance and stability. DCD uses counterfactual probability to formulate a distributional explanatory relation, not to analyze the truth conditions or probabilities of individual counterfactual sentences.

### 4.3 Distributional Practices

The special sciences already contain distributional practices: response-time models, wage-decomposition procedures, and estimators of selection gradients all measure, model, identify, and compare distributions. DCD asks how such practices bear on the metaphysical question of non-redundant higher-level explanation under lower-level dependence[5].

In psychology, response-time distribution analyses show that means may obscure theoretically relevant distributional changes (Balota & Yap, 2011; Heathcote et al., 1991). In labor economics, counterfactual density methods, unconditional quantile regression, and distribution-regression approaches study wage distributions and their functionals (DiNardo et al., 1996; Firpo et al., 2009; Fortin et al., 2021). In evolutionary biology, selection-gradient analysis and flexible fitness-function modeling relate phenotypic variation to fitness and population-level change (Lande & Arnold, 1983; Morrissey & Sakrejda, 2013). DCD does not supply these methods; it interprets their relevance to the autonomy debate.

The philosophical issue concerns the explanatory role of the distributional mapping. Conditions can be mapped to distributions in many ways, while robustness, identification, sensitivity analysis, and model selection remain empirical matters. In the notation of Section 3, $\Psi_Y$ captures the dependence invoked by a higher-level explanation only when $T$ tracks the relevant distributional feature and the contrast remains stable across $C^*$ at the relevant grain.

[5] The appeal to counterfactual distributions is philosophical rather than methodological. Potential-outcome and causal-modeling frameworks already provide tools for representing and identifying such quantities (Holland, 1986; Pearl, 2009a, 2009b; Rubin, 1974). The question here is when those distributions, once treated as explananda, are preserved under lower-level reconstruction.

### 4.4 Replacement Without Loss

If higher-level properties depend on lower-level bases, higher-level autonomy cannot consist in independence from those bases. But dependence alone does not establish redundancy. Redundancy requires a lower-level description that preserves the distributional structure relevant to the explanation. This point aligns with accounts on which nonfundamental explanations may gain depth by abstracting away from lower-level detail (Weslake, 2010).

Given lower-level dependence, $\Psi_Y$ may in principle be recovered from lower-level bases and warranted relations between the variables. Whether this amounts to replacement depends on what the reconstruction preserves. A lower-level reconstruction specified independently of the distribution to be explained replaces $\Psi_Y$ only when it preserves the explanandum and contrast, the relevant distributional feature, the background range, and the explanatory grain[6]. Preserving grain does not require retaining higher-level vocabulary. In the hypothetical example in subsection 3.2.2, the neural transmission rule explains the load-related distributional change, whereas the second reconstruction matches the marginal distributions but loses their joint organization.

Building on work on proportionality, realization-insensitivity, and different types of generality across levels of explanation, DCD locates the gap between prediction and explanation within the relation between two distribution-valued dependence structures (Menzies & List, 2010; Potochnik, 2010; Woodward, 2021b; Yablo, 1992). In Section 3's terms, $\Psi_Y$ may depend on structures governed by $\Phi$ without becoming explanatorily dispensable. Dispensing with $\Psi_Y$ requires preserving that explanatory role, not merely descending to lower-level bases.

Kim's challenge thus becomes a question about what a lower-level explanation preserves. For a distributional explanandum, redundancy requires recovering the relevant counterfactual pattern from lower-level processes at the same explanatory grain. DCD locates higher-level explanatory significance in this pattern and allows lower-level replacement when the same explanatory role is preserved.

## 5 Conclusion

The autonomy of the special sciences can rest on the organization of explanatory dependence across levels, even when higher-level facts depend on lower-level bases. Explanatorily significant DCD is one form of this organization: a higher-level contrast makes a stable counterfactual difference to those features of an outcome distribution that matter to the question at hand. A lower-level model replaces such an explanation without loss only if it is specified independently of the distribution to be explained and preserves the explanandum and contrast, the relevant distributional feature, the background range, and the explanatory grain. Which models meet these conditions depends on the empirical details of the case.

### Acknowledgments

I thank Alan Hájek for reading an early draft and raising questions that exposed central problems in the original formulation of DCD. I also thank the anonymous reviewers for comments that helped clarify the argument and improve its presentation.

### Declarations

**Disclosure of Interests:** The author declares no competing interests.
**Funding:** This study was supported by the Major Program of the National Fund of Philosophy and Social Science of China (Grant No. 22&ZD046) and the China Scholarship Council (Grant No. 202606070045).
**Ethical approval:** Not applicable.
**Informed consent:** Not applicable.
**Author contribution:** Youheng Zhang is the sole author of this manuscript.
**Data Availability Statement:** Not applicable.

[6] For related discussion of explanation by idealization and asymptotic reasoning, see Batterman (2001, 2010). In those cases too, retaining more lower-level detail is not automatically an explanatory improvement.

## References

Abrams, M. (2006). Infinite populations and counterfactual frequencies in evolutionary theory. *Studies in History and Philosophy of Science Part C: Studies in History and Philosophy of Biological and Biomedical Sciences, 37*(2), 256–268. https://doi.org/10.1016/j.shpsc.2006.03.004

Bacciagaluppi, G. (2020). Unscrambling subjective and epistemic probabilities. In M. Hemmo & O. Shenker (Eds.), *Quantum, probability, logic: The work and influence of Itamar Pitowsky* (pp. 49–89). Springer International Publishing. https://doi.org/10.1007/978-3-030-34316-3_3

Balke, A., & Pearl, J. (1994). Counterfactual probabilities: Computational methods, bounds and applications. In R. L. de Mantaras & D. Poole (Eds.), *Uncertainty in artificial intelligence* (pp. 46–54). Morgan Kaufmann. https://doi.org/10.1016/B978-1-55860-332-5.50011-0

Balota, D. A., & Yap, M. J. (2011). Moving beyond the mean in studies of mental chronometry: The power of response time distributional analyses. *Current Directions in Psychological Science, 20*(3), 160–166. https://doi.org/10.1177/0963721411408885

Barker, S. (1999). Counterfactuals, probabilistic counterfactuals and causation. *Mind, 108*(431), 427–469. https://doi.org/10.1093/mind/108.431.427

Batterman, R. W. (2001). Asymptotic reasoning. In R. W. Batterman, *The devil in the details: Asymptotic reasoning in explanation, reduction, and emergence* (pp. 9–12). Oxford University Press. https://doi.org/10.1093/0195146476.003.0002

Batterman, R. W. (2010). On the explanatory role of mathematics in empirical science. *The British Journal for the Philosophy of Science, 61*(1), 1–25. https://doi.org/10.1093/bjps/axp018

Bennett, K. (2003). Why the exclusion problem seems intractable, and how, just maybe, to tract it. *Noûs, 37*(3), 471–497.

Chernozhukov, V., Fernández-Val, I., & Melly, B. (2013). Inference on counterfactual distributions. *Econometrica, 81*(6), 2205–2268. https://doi.org/10.3982/ECTA10582

Davis, W. A. (1980). Swain's counterfactual analysis of causation. *Philosophical Studies, 38*(2), 169–176. https://doi.org/10.1007/BF00374333

DiNardo, J., Fortin, N. M., & Lemieux, T. (1996). Labor market institutions and the distribution of wages, 1973–1992: A semiparametric approach. *Econometrica, 64*(5), 1001–1044. https://doi.org/10.2307/2171954

Engle, R. W. (2002). Working memory capacity as executive attention. *Current Directions in Psychological Science, 11*(1), 19–23. https://doi.org/10.1111/1467-8721.00160

Finch, H. A. (1958). An explication of counterfactuals by probability theory. *Philosophy and Phenomenological Research, 18*(3), 368–378. https://doi.org/10.2307/2104541

Firpo, S., Fortin, N. M., & Lemieux, T. (2009). Unconditional quantile regressions. *Econometrica, 77*(3), 953–973. https://doi.org/10.3982/ECTA6822

Fodor, J. A. (1974). Special sciences, or: The disunity of science as a working hypothesis. *Synthese, 28*(2), 97–115. https://doi.org/10.1007/BF00485230

Fodor, J. A. (1997). Special sciences: Still autonomous after all these years. *Noûs, 31*(S11), 149–163. https://doi.org/10.1111/0029-4624.31.s11.7

Fortin, N., Lemieux, T., & Firpo, S. (2011). Decomposition methods in economics. In O. Ashenfelter & D. Card (Eds.), *Handbook of labor economics* (Vol. 4, pp. 1–102). Elsevier. https://doi.org/10.1016/S0169-7218(11)00407-2

Fortin, N. M., Lemieux, T., & Lloyd, N. (2021). Labor market institutions and the distribution of wages: The role of spillover effects. *Journal of Labor Economics, 39*(S2), S369–S412. https://doi.org/10.1086/712923

Glennan, S. (2009). Productivity, relevance and natural selection. *Biology & Philosophy, 24*(3), 325–339. https://doi.org/10.1007/s10539-008-9137-7

Hájek, A. (2003). What conditional probability could not be. *Synthese, 137*(3), 273–323.

Hájek, A. (2011). Triviality pursuit. *Topoi, 30*(1), 3–15. https://doi.org/10.1007/s11245-010-9083-2

Hájek, A. (2012). The fall of "Adams' thesis"? *Journal of Logic, Language and Information, 21*(2), 145–161. https://doi.org/10.1007/s10849-012-9157-1

Hájek, A. (2014). *Most counterfactuals are false*. PhilArchive. https://philarchive.org/rec/HJEMCA

Hájek, A. (2021a). Contra counterfactism. *Synthese, 199*(1), 181–210. https://doi.org/10.1007/s11229-020-02643-x

Hájek, A. (2021b). Counterfactual scepticism and antecedent-contextualism. *Synthese, 199*(1), 637–659. https://doi.org/10.1007/s11229-020-02686-0

Hájek, A. (2025). Similarity accounts of counterfactuals: A reality check. *Philosophy and Phenomenological Research, 110*(3), 887–915. https://doi.org/10.1111/phpr.13138

Heathcote, A., Popiel, S. J., & Mewhort, D. J. (1991). Analysis of response time distributions: An example using the Stroop task. *Psychological Bulletin, 109*(2), 340–347. https://doi.org/10.1037/0033-2909.109.2.340

Hédoin, C. (2019). Institutions, rule-following and conditional reasoning. *Journal of Institutional Economics, 15*(1), 1–25. https://doi.org/10.1017/S1744137418000073

Hoefer, C. (2007). The third way on objective probability: A sceptic's guide to objective chance. *Mind, 116*(463), 549–596. https://doi.org/10.1093/mind/fzm549

Holland, P. W. (1986). Statistics and causal inference. *Journal of the American Statistical Association, 81*(396), 945–960. https://doi.org/10.1080/01621459.1986.10478354

Huneman, P. (2012). Natural selection: A case for the counterfactual approach. *Erkenntnis, 76*(2), 171–194. https://doi.org/10.1007/s10670-011-9306-y

Kallestrup, J. (2006). The causal exclusion argument. *Philosophical Studies, 131*(2), 459–485. https://doi.org/10.1007/s11098-005-1439-x

Kim, J. (1973). Causes and counterfactuals. *The Journal of Philosophy, 70*(17), 570–572. https://doi.org/10.2307/2025312

Kim, J. (1974). Noncausal connections. *Noûs, 8*(1), 41–52. https://doi.org/10.2307/2214644

Kim, J. (1992). Multiple realization and the metaphysics of reduction. *Philosophy and Phenomenological Research, 52*(1), 1–26. https://doi.org/10.2307/2107741

Kim, J. (1995). Mental causation: What? Me worry? *Philosophical Issues, 6*, 123–151. https://doi.org/10.2307/1523036

Kim, J. (2003). Blocking causal drainage and other maintenance chores with mental causation. *Philosophy and Phenomenological Research, 67*(1), 151–176.

Kim, J. (2006). Emergence: Core ideas and issues. *Synthese, 151*(3), 547–559. https://doi.org/10.1007/s11229-006-9025-0

Kim, J. (2007). The supervenience argument motivated, clarified, and defended. In *Physicalism, or something near enough* (pp. 32–92). Princeton University Press. https://doi.org/10.1515/9781400840847.32

Kim, J. (2010). Explanatory knowledge and metaphysical dependence. In J. Kim, *Essays in the metaphysics of mind* (pp. 167–186). Oxford University Press. https://doi.org/10.1093/acprof:oso/9780199585878.003.0009

Kim, J. (2013). "Supervenient and yet not deducible": Is there a coherent concept of ontological emergence? In A. Hieke & H. Leitgeb (Eds.), *Reduction: Between the mind and the brain* (pp. 53–72). De Gruyter. https://doi.org/10.1515/9783110328851.53

Kim, J. (2014). Supervenience and causal closure: Two exclusion arguments and how in the end they reduce to one. *철학, 118*, 1–25.

Kim, J. (2023). Causation and mental causation. In B. McLaughlin & J. Cohen (Eds.), *Contemporary debates in philosophy of mind* (pp. 215–229). John Wiley & Sons. https://doi.org/10.1002/9781394259847.ch13

Lande, R., & Arnold, S. J. (1983). The measurement of selection on correlated characters. *Evolution, 37*(6), 1210–1226. https://doi.org/10.1111/j.1558-5646.1983.tb00236.x

Leitgeb, H. (2012). A probabilistic semantics for counterfactuals. Part A. *The Review of Symbolic Logic, 5*(1), 26–84. https://doi.org/10.1017/S1755020311000153

Lewis, D. (1973). Causation. *The Journal of Philosophy, 70*(17), 556–567. https://doi.org/10.2307/2025310

Lewis, D. (1979). Counterfactual dependence and time's arrow. *Noûs, 13*(4), 455–476. https://doi.org/10.2307/2215339

Lewis, D. (2000). Causation as influence. *The Journal of Philosophy, 97*(4), 182–197. https://doi.org/10.2307/2678389

List, C., & Pivato, M. (2021). Dynamic and stochastic systems as a framework for metaphysics and the philosophy of science. *Synthese, 198*(3), 2551–2612. https://doi.org/10.1007/s11229-019-02231-8

Lyon, A. (2011). Deterministic probability: Neither chance nor credence. *Synthese, 182*(3), 413–432. https://doi.org/10.1007/s11229-010-9750-2

Ma, S., Yan, X., Ma, L., Billington, J., Merat, N., & Markkula, G. (2026). Modelling cognitive load using drift-diffusion models in pedestrian street-crossing: A method supported by neural evidence. *Accident Analysis & Prevention, 225*, 108310. https://doi.org/10.1016/j.aap.2025.108310

Menzies, P., & List, C. (2010). The causal autonomy of the special sciences. In C. Macdonald & G. Macdonald (Eds.), *Emergence in mind* (pp. 108–128). Oxford University Press. https://doi.org/10.1093/acprof:oso/9780199583621.003.0008

Meyering, T. C. (2000). Physicalism and downward causation in psychology and the special sciences. *Inquiry, 43*(2), 181–202. https://doi.org/10.1080/002017400407744

Millstein, R. L. (2006). Natural selection as a population-level causal process. *The British Journal for the Philosophy of Science, 57*(4), 627–653. https://doi.org/10.1093/bjps/axl025

Morrissey, M. B., & Sakrejda, K. (2013). Unification of regression-based methods for the analysis of natural selection. *Evolution, 67*(7), 2094–2100. https://doi.org/10.1111/evo.12077

Muhlert, M. (2025). *Cognitive immunology: The adaptive falsification hypothesis and the evolutionary origins of scientific skepticism* (SSRN Scholarly Paper No. 5409502). Social Science Research Network. https://doi.org/10.2139/ssrn.5409502

Oulasvirta, A., & Hornbæk, K. (2022). Counterfactual thinking: What theories do in design. *International Journal of Human–Computer Interaction, 38*(1), 78–92. https://doi.org/10.1080/10447318.2021.1925436

Pearl, J. (1999). Probabilities of causation: Three counterfactual interpretations and their identification. *Synthese, 121*(1), 93–149. https://doi.org/10.1023/A:1005233831499

Pearl, J. (2009a). Causal diagrams and the identification of causal effects. In *Causality* (2nd ed., pp. 65–106). Cambridge University Press. https://doi.org/10.1017/CBO9780511803161.005

Pearl, J. (2009b). The logic of structure-based counterfactuals. In *Causality* (2nd ed., pp. 201–258). Cambridge University Press. https://doi.org/10.1017/CBO9780511803161.009

Pietsch, W. (2022). Causal probability. In W. Pietsch (Ed.), *On the epistemology of data science: Conceptual tools for a new inductivism* (pp. 235–287). Springer International Publishing. https://doi.org/10.1007/978-3-030-86442-2_9

Potochnik, A. (2010). Levels of explanation reconceived. *Philosophy of Science, 77*(1), 59–72. https://doi.org/10.1086/650208

Potochnik, A. (2021). Our world isn't organized into levels. In D. S. Brooks, J. DiFrisco, & W. C. Wimsatt (Eds.), L*evels of organization in the biological sciences* (pp. 61–76). MIT Press. https://doi.org/10.7551/mitpress/12389.003.0007

Qian, L. (2025). Weak-necessity causal reasoning for evaluating counterfactual arguments in law and economics. *Journal of Economic Methodology, 32*(4), 301–316. https://doi.org/10.1080/1350178X.2025.2589110

Raatikainen, P. (2010). Causation, exclusion, and the special sciences. *Erkenntnis, 73*(3), 349–363. https://doi.org/10.1007/s10670-010-9236-0

Ratcliff, R., & McKoon, G. (2008). The diffusion decision model: Theory and data for two-choice decision tasks. *Neural Computation, 20*(4), 873–922. https://doi.org/10.1162/neco.2008.12-06-420

Rescorla, M. (2018). An interventionist approach to psychological explanation. *Synthese, 195*(5), 1909–1940. https://doi.org/10.1007/s11229-017-1553-2

Reutlinger, A. (2013). Causation in the special sciences. In A. Reutlinger (Ed.), *A theory of causation in the social and biological sciences* (pp. 3–24). Palgrave Macmillan. https://doi.org/10.1057/9781137281043_1

Rubin, D. B. (1974). Estimating causal effects of treatments in randomized and nonrandomized studies. *Journal of Educational Psychology, 66*(5), 688–701. https://doi.org/10.1037/h0037350

Sandgren, A., & Steele, K. (2021). Levelling counterfactual scepticism. *Synthese, 199*(1), 927–947. https://doi.org/10.1007/s11229-020-02742-9

Santorio, P. (2025). Probabilities of counterfactuals are counterfactual probabilities. *The Journal of Philosophy, 122*(6), 285–312. https://doi.org/10.5840/jphil2025122613

Schrenk, M. (2006). *A theory for special science laws*. PhilArchive. https://philarchive.org/rec/SCHATF

Shapiro, L. A. (2000). Multiple realizations. *The Journal of Philosophy, 97*(12), 635–654. https://doi.org/10.2307/2678460

Sober, E. (1999). The multiple realizability argument against reductionism. *Philosophy of Science, 66*(4), 542–564. https://doi.org/10.1086/392754

Sweller, J. (1988). Cognitive load during problem solving: Effects on learning. *Cognitive Science, 12*(2), 257–285. https://doi.org/10.1207/s15516709cog1202_4

Weslake, B. (2010). Explanatory depth. *Philosophy of Science*, *77*(2), 273–294. https://doi.org/10.1086/651316

Woodward, J. (2000). Explanation and invariance in the special sciences. *The British Journal for the Philosophy of Science, 51*(2), 197–254. https://doi.org/10.1093/bjps/51.2.197

Woodward, J. (2004). Invariance. In *Making things happen: A theory of causal explanation* (pp. 239–314). Oxford University Press. https://doi.org/10.1093/0195155270.003.0006

Woodward, J. (2021a). Explanatory autonomy: The role of proportionality, stability, and conditional irrelevance. *Synthese, 198*(1), 237–265. https://doi.org/10.1007/s11229-018-01998-6

Woodward, J. (2021b). Proportionality. In *Causation with a human face: Normative theory and descriptive psychology* (pp. 357–392). Oxford University Press. https://doi.org/10.1093/oso/9780197585412.003.0009

Yablo, S. (1992). Mental causation. *The Philosophical Review, 101*(2), 245–280. https://doi.org/10.2307/2185535

Zhang, Y. (2026). Stochastic supervenience: A conservative extension of classical deterministic supervenience. *European Journal for Philosophy of Science, 16*(2), Article 32. https://doi.org/10.1007/s13194-026-00739-8